\providecommand\reserveinserts[1]{}
\documentclass[AMA,Times1COL]{WileyNJDv5} 

\articletype{Research Article - Empirical}%

\received{Date Month Year}
\revised{Date Month Year}
\accepted{Date Month Year}
\journal{Journal}
\volume{00}
\copyyear{2023}
\startpage{1}

\usepackage{pifont}
\usepackage{array}
\usepackage{adjustbox}
\usepackage{wrapfig}
\usepackage{threeparttable}
\usepackage{threeparttablex} 
\usepackage{graphicx}
\usepackage[figuresright]{rotating}
\usepackage[misc]{ifsym} 
\usepackage{hyperref}
\usepackage{tcolorbox}
\usepackage{booktabs}
\DeclareRobustCommand{\lqy}[1]{#1}

\definecolor{background}{rgb}{0.94, 0.97, 1.0}
\definecolor{edge}{rgb}{0.32, 0.48, 0.72}

\makeatletter
\newcommand{\thickhline}{%
    \noalign {\ifnum 0=`}\fi \hrule height 1pt
    \futurelet \reserved@a \@xhline
}

\begin{document}

\title{Towards the Generalizability of Leveraging ChatGPT in APR via Self-enhancing: An Empirical Study}

\author[1]{Qingyuan Li}
\author[1]{Chuanyi Li}
\author[1]{Yaopeng Yang}
\author[1]{Ziwen Ge}
\author[1]{Jidong Ge}
\author[1]{Bin Luo}

\authormark{Li \textsc{et al.}}
\titlemark{Towards the Generalizability of Leveraging ChatGPT in APR via Self-enhancing: An Empirical Study}

\address[1]{\orgdiv{State Key Laboratory for Novel Software Technology}, \orgname{Nanjing University}, \orgaddress{\state{Jiangsu}, \country{China}}}

\corres{Chuanyi Li (lcy@nju.edu.cn)}



\abstract[Abstract]{Recent work on Automated Program Repair (APR) has increasingly shifted its foundation from fine-tuned pre-trained code models to Large Language Models (LLMs). Against this backdrop, a growing body of work, referred to as ChatGPT-enhanced APR, leverages advanced techniques (e.g., self-correction and autonomous agents) to improve ChatGPT’s repair capability without modifying model parameters. Although these approaches report strong results on challenging benchmarks such as Defects4J and SWE-bench, whether they deliver stable enhancement gains across different benchmarks remains under-explored. To assess the generalizability of ChatGPT-enhanced APR across benchmarks, we conduct an empirical study evaluating three ChatGPT-enhanced APRs on three representative and long-standing benchmarks. 
\lqy{ 
Experimental results show that enhancement gains vary across benchmarks. With GPT-3.5-Turbo, SRepair achieves a larger absolute gain on HumanEval-Java than on Defects4J, while SRepair and FixAgent without extrinsic information yield negative gains on BugsInPy. With GPT-5.4-mini, the evaluated methods achieve larger absolute gains on Defects4J than on HumanEval-Java, while gains on BugsInPy are non-negative but remain limited.
We subsequently investigate benchmark-related factors that may affect enhancement gains through code transformations and benchmark-specific fine-tuning. \lqy{Code transformations reduce enhancement gains on Defects4J, while benchmark-specific fine-tuning increases the gains on BugsInPy.} Furthermore, directly supplying GPT-3.5-Turbo with error messages and triggering tests yields more correct repairs than the evaluated ChatGPT-enhanced APR methods on BugsInPy. Taken together, these findings highlight the need to evaluate the generalizability of ChatGPT-enhanced APR across benchmarks and models using multiple metrics, and suggest that directly providing repair-specific extrinsic information may be more effective than enhancement methods when their gains are limited.
}

}



\maketitle

\renewcommand\thefootnote{}

\renewcommand\thefootnote{\fnsymbol{footnote}}
\setcounter{footnote}{1}


\section{Introduction} \label{section:I}

The increasing scale and complexity of software systems have led to a growing number and diversity of defects, posing severe threats to software security and society \cite{telang2007empirical}. Nonetheless, repairing software defects is a highly labor-intensive and costly process \cite{gazzola2018automatic}, where Automated Program Repair (APR) aims to alleviate this burden by automatically generating patches to fix defects \cite{monperrus2018automatic}. 
Recently, Large Language Models (LLMs) have been widely adopted in software engineering \cite{gao2025current,ramos2025ai,nejjar2025llms} and have demonstrated notable progress on APR \cite{machavcek2025impact,li2024exploring,huang2023empirical,silva2025repairllama}.
Beyond direct prompting, a range of LLM-enhanced approaches, such as retrieval-augmented generation (RAG) \cite{gao2024retrieval}, self-correction \cite{madaan2024self, miao2023selfcheck}, and agent-based workflows \cite{wang2024survey}, have been extensively explored for program repair. 
Meanwhile, the target scope of repair has shifted from single functions or files toward repository-level defects. Along with this shift, recent evaluations in LLM-based APR have increasingly emphasized the SWE-bench family (e.g., SWE-bench \cite{jimenez2024swebench} and SWE-bench Live \cite{zhang2025swe}), while established benchmarks are less frequently revisited in contemporary studies. While LLM-based APRs achieve strong performance on the SWE-bench family, long-standing benchmarks remain far from solved. For example, the state-of-the-art method repairs only 50\% of defects on BugsInPy \cite{ehsani2025bug}. 
Consequently, these classic benchmarks remain valuable for rigorously evaluating modern LLM-based APRs and for characterizing their ability to generalize beyond the SWE-bench family.

Among these LLM-based program repair tools, a class of APR approaches known as ChatGPT-enhanced APR leverages LLM-specific enhancement strategies (e.g., self-correction \cite{madaan2024self}, generated-knowledge prompting \cite{liu2022generated, huang2025empowering}, and autonomous agents \cite{wang2024survey}) to better elicit and use ChatGPT’s inherent knowledge~\footnote{The inherent knowledge of an LLM is encoded in its model parameters \cite{huanglarge}.}, without modifying model parameters. 
For example, TRP \cite{yu2024fight} enhances repair via self-correction by incorporating test-related feedback into the repair loop; SRepair \cite{xiang2024far} uses generated-knowledge prompting to produce root-cause and repair suggestions to guide patch generation; and FixAgent \cite{lee2024unified} adopts a multi-agent workflow that decomposes debugging into specialized roles and iteratively aggregates their outputs.
A key difference among ChatGPT-enhanced APR approaches is how they use repair-specific information \cite{yang2025patch}: many approaches rely on \emph{extrinsic defect information} to enrich bug context, whereas only a few rely solely on \emph{intrinsic defect information}.
We introduce this distinction because we observe that enhancement gains can be unstable across benchmarks, and different enhancement methods both (1) rely on different types of extrinsic information and (2) may face different extrinsic information availability across benchmarks and real-world settings. In real-world scenarios, extrinsic information is not always provided to the repair tool, as obtaining it incurs an additional cost. For example, a significant portion of buggy code lacks relevant tests, as it would take a developer considerable time to write tests and collect corresponding error messages.
To understand the generalizability of ChatGPT-enhanced APRs, it is necessary to consider the gains brought by different extrinsic information and to isolate the generalizability of self-enhancement when extrinsic information is absent.
\lqy{Specifically, intrinsic information refers to the \textit{buggy context}, namely the code to be repaired. The given method-level fault location is a separate experimental condition shared by all evaluated methods.} Extrinsic information refers to information obtained by relying on artifacts beyond the defect itself, such as developers’ annotations about the bug (i.e., \textit{bug comments}), \textit{test cases} that trigger the defect, or \textit{error messages} and \textit{debugging information} collected during dynamic execution of the buggy code. Table~\ref{tab:extrinsicinformation} shows the extrinsic information used by different ChatGPT-enhanced APRs. 
\begin{table}[htbp]
\small
\centering
\caption{Extrinsic information used by ChatGPT-enhanced APRs}
\begin{adjustbox}{max width=0.8\columnwidth}
\begin{tabular}{cccccc}
\toprule
\multirow{2}*{\textbf{Method}} & \multicolumn{5}{c}{\textbf{Extrinsic Information}} \\
\cmidrule(lr){2-6}
\textbf{} & \textit{Error Message} & \textit{Test} & \textit{Bug Comment} & \textit{Code Description} & \textit{DebugInfo} \\
\cmidrule{1-6}\morecmidrules\cmidrule{1-6}

\textit{FixAgent} \cite{lee2024unified} & \textcolor{green}{\ding{51}} & \textcolor{green}{\ding{51}} & \textcolor{red}{\ding{55}} & \textcolor{green}{\ding{51}} & \textcolor{green}{\ding{51}} \\
\textit{RepairAgent} \cite{bouzenia2024repairagent} & \textcolor{green}{\ding{51}} & \textcolor{green}{\ding{51}} & \textcolor{red}{\ding{55}} & \textcolor{red}{\ding{55}} & \textcolor{green}{\ding{51}} \\
\textit{SRepair} \cite{xiang2024far} & \textcolor{green}{\ding{51}} & \textcolor{green}{\ding{51}} & \textcolor{green}{\ding{51}} & \textcolor{red}{\ding{55}} & \textcolor{red}{\ding{55}} \\
\textit{TRP} \cite{yu2024fight} & \textcolor{red}{\ding{55}} & \textcolor{red}{\ding{55}} & \textcolor{red}{\ding{55}} & \textcolor{green}{\ding{51}} & \textcolor{red}{\ding{55}} \\
\bottomrule
\end{tabular}\label{tab:extrinsicinformation}
\end{adjustbox}
\footnotesize
\end{table}



Despite ChatGPT-enhanced APRs achieving strong results on challenging benchmarks, existing evaluations typically consider only one or two benchmarks. For example, SRepair \cite{xiang2024far} and RepairAgent \cite{bouzenia2024repairagent} demonstrate strong repair performance on Defects4J \cite{just2014defects4j}, repairing 300 and 164 defects, respectively; however, their evaluation are limited to Defects4J. REFINE \cite{pabba2025refineenhancingprogramrepair} achieves resolution rates of 51.67\% on SWE-Bench Lite and 63.8\% on SWE-Bench Verified; however, it is only evaluated on the SWE-Bench family.
Even with these successes, widely used benchmarks remain under-explored, such as HumanEval-Java \cite{jiang2023impact}, BugsInPy \cite{widyasari2020bugsinpy}, and TypeBugs \cite{oh2022pyter}, leaving the generalizability of these ChatGPT-enhanced APRs unclear. This limitation raises our first question: Can ChatGPT-enhanced APRs generalize across a broader range of benchmarks? To address this question, we systematically evaluate the enhancement effectiveness of three ChatGPT-enhanced APR approaches across three representative and long-standing benchmarks, including Defects4J, HumanEval-Java, and BugsInPy.

\lqy{The results show that enhancement gains depend on the benchmark, model, and extrinsic information. 
With GPT-3.5-Turbo, SRepair and FixAgent achieve non-negative gains on all three benchmarks, whereas RefinedTRP, which uses no extrinsic information, generally yields smaller gains. Specifically, SRepair improves the repair rate by 38.65 percentage points (pp) on HumanEval-Java and 10.06 pp on Defects4J, despite a lower performance ratio (PIR) on HumanEval-Java, and provides no gain on BugsInPy. FixAgent achieves absolute gains of 10.43 pp, 8.74 pp, and 3.70 pp on HumanEval-Java, Defects4J, and BugsInPy, respectively. The corresponding gains for RefinedTRP are 9.82 pp, 1.08 pp, and 0.00 pp. However, removing extrinsic information reduces the gains of SRepair and FixAgent on all three benchmarks, yielding negative net gains of $-3.70$ pp and $-1.85$ pp, respectively, on BugsInPy.
With GPT-5.4-mini, enhancement gains also vary across benchmarks. The evaluated methods achieve larger absolute gains on Defects4J than on HumanEval-Java. However, their net headroom gains reach 100\% on HumanEval-Java, compared with 3.56\% to 7.55\% on Defects4J, because all three methods repair the three defects remaining unresolved by the baseline. Their PIRs are closer to parity: 101.88\% on HumanEval-Java and 104.75\% to 110.06\% on Defects4J. Gains on BugsInPy remain limited despite being non-negative. These results highlight the need to evaluate the generalizability of ChatGPT-enhanced APR across benchmarks and models using multiple metrics.
}
These results naturally motivate our second question: what factors lead to the varying enhancement ratios of ChatGPT-enhanced APRs across different benchmarks? Zhou et al. \cite{zhou2023don} show that benchmarks commonly used are more likely to be included in LLM training datasets due to data leakage, increasing the likelihood that these models memorize them and leading to enhanced performance on these benchmarks. \lqy{Inspired by this work, we measure benchmark memorization indicators and use them to motivate exploratory experiments involving code transformations and benchmark-specific adaptation. \lqy{These experiments examine how enhancement gains respond to changes in input form and benchmark-specific adaptation.}}

To present the detailed experimental process and findings of the aforementioned study on the generalizability of ChatGPT-enhanced APRs, we structured our empirical study around four research questions (RQs): 
\textit{\textbf{RQ1}-How do ChatGPT-enhanced APRs exhibit generalizability?},
\textit{\textbf{RQ2}-How well does ChatGPT memorize APR benchmarks?},
\lqy{\textit{\textbf{RQ3}-How do enhancement gains change under code transformations and benchmark-specific adaptation motivated by memorization indicators?}}, and
\textit{\textbf{RQ4}-How effective is directly providing extrinsic information to ChatGPT on benchmarks with low memorization?}

\lqy{Our study yields four findings: \textbf{Enhancement gains vary across benchmarks, model configurations, and extrinsic information}, and their interpretation requires both absolute and baseline-relative metrics. \textbf{ChatGPT’s memorization varies across different benchmarks}, with stronger clone-matching signals for Defects4J than for BugsInPy. \textbf{Enhancement gains change under code transformations and benchmark-specific adaptation}, \lqy{with smaller gains under code transformations and method-dependent changes after adaptation.} \textbf{Repair-specific extrinsic information helps in the evaluated BugsInPy}: error messages together with triggering tests improve GPT-3.5-Turbo's results on the 54 BugsInPy type-related defects.}

\lqy{Our empirical study suggests three directions for future research: (1) evaluate enhancement on multiple benchmarks using repair rates and complementary gain metrics; (2) investigate which repair-specific information benefits each workflow; and (3) \lqy{evaluate how benchmark-specific adaptation affects general repair ability, using memorization indicators to inform the analysis.}}

\section{Related Work}
\label{section:II}

\subsection{Strategies of Applying LLMs to SE Tasks}
Large Language Models \cite{zhao2023survey,chang2024survey,zan2023large} represent one of the most revolutionary advancements in deep learning (DL) \cite{pouyanfar2018survey,dong2021survey} and natural language processing (NLP) \cite{goldberg2016primer,chowdhary2020natural}. In recent years, LLMs have been increasingly applied to downstream tasks in software engineering \cite{wang2025exploring,diebold2025backlogs,ramos2025ai,nejjar2025llms,talha2025semiautomated}. When utilized for code-related tasks, LLMs are typically employed in the following ways: 
(1) \textit{Training-based methods}: Pre-training or fine-tuning is the standard way to impart LLMs with code-related knowledge. CodeLlama \cite{roziere2023code} is a code model pre-trained on 500 billion code snippets, enabling Llama2 \cite{touvron2023llama} to generalize code knowledge. Building upon CodeLlama, Jiang et al. \cite{jiang2024training} implement supervised fine-tuning (SFT) \cite{chung2024scaling,ding2023parameter}, enhancing CodeLlama's capability for self-debugging in code generation.
(2) \textit{Domain knowledge augmentation}: Training alone may sometimes fail to provide LLMs with domain-specific knowledge, such as repository-level information. To overcome this, techniques like Retrieval Augmented Generation (RAG) and KNN-LM \cite{khandelwal2020nearest} often provide LLMs with more domain-specific knowledge. InferFix \cite{jin2023inferfix} improves repair performance by prompting LLM with the context containing similar fixes. KNM-LM \cite{tang2023domain} employs a KNN-based method that integrates domain database knowledge with LLM inference results, significantly boosting code completion.

\subsection{Techniques of Enhancing Large Language Models}
The LLM enhancement technique is emerging in the field of natural language processing. A common characteristic of these techniques is that they enable LLMs to perform additional operations while conducting inference, such as forming CoT, generating additional content to aid reasoning, self-correcting the inference process or results, or acting as autonomous agents (for example, assuming specific roles for critiquing or collaborating). \lqy{LLM enhancement workflows typically combine several techniques.} For instance, when utilizing self-correction, a CoT prompt is often used to guide LLMs to adopt multi-step reasoning. This approach decomposes complex tasks, allowing LLMs to self-check the intermediate steps of the reasoning process \cite{miao2023selfcheck}; on the other hand, models may self-refine their reasoning results iteratively through feedback \cite{madaan2024self}. While employing agent-based techniques, multiple agents are usually involved and collaborate to solve problems together \cite{lee2024unified}, if all agents are treated as a united system, this approach resembles a CoT utilizing multi-step reasoning; another approach involving multiple agents drives them to critique and debate each other's generated content, leading to enhanced performance \cite{du2023improving,liang2023encouraging}, if all agents are considered as a united system, this approach is similar to a system engaging in self-correction.

\subsection{Memorization in Large Language Models}
In deep learning, a model's ability can be generally interpreted as both memorization and generalization \cite{elangovan2021memorization}. Memorization in LLMs is often defined as their ability to recall and reproduce \lqy{the data used to train them}. In the context of generation tasks, Yang et al. \cite{yang2024unveiling} define the memorization fragments of LLMs as strings that appear in both the training dataset and can be generated by the LLMs based on a given prompt. They discuss the security and privacy issues of memorization in code models from the perspectives of intellectual property, security vulnerabilities, and sensitive information leakage. Furthermore, they design two methods to detect memorization in LLMs depending on the accessibility of the training data. Memorization Detection, applicable when training data is accessible, involves a sliding window approach to count the number of Type-1 clones \cite{roy2007survey} between the generated content and the training data within the window, thus assessing the degree of memorization. Memorization Prediction, on the other hand, used when training data is inaccessible, evaluates the memorization degree by calculating metrics of generated content such as Perplexity, PPL-PPL ratio, PPL-zlib ratio, and Average PPL \cite{carlini2021extracting}.

\section{Study Design}
\label{section:III}
\begin{figure*}[htbp]
    \centering
        {\includegraphics[width=0.9\linewidth]{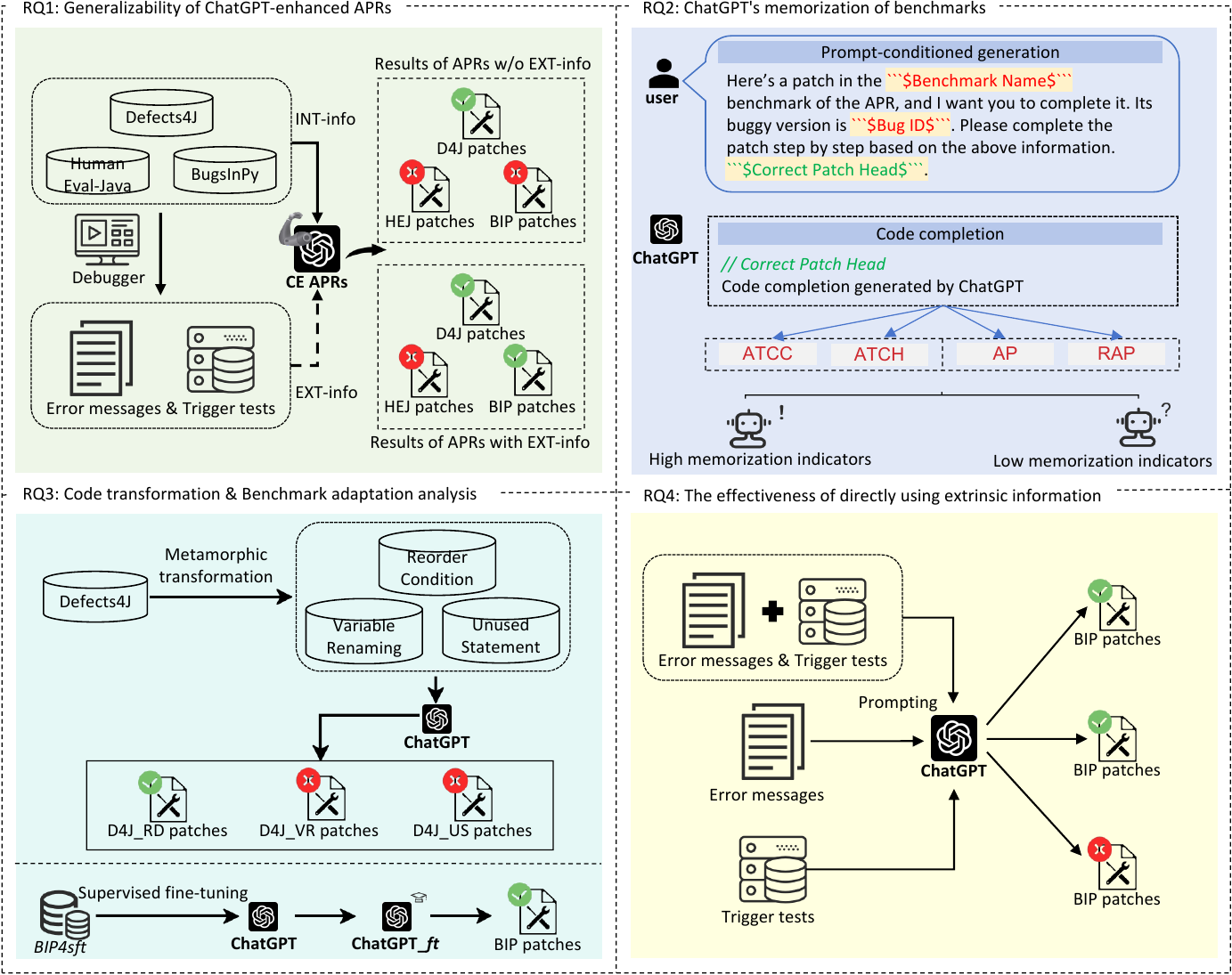}}
    \caption{Overview of the empirical study.}
    \label{fig:figure1}
\end{figure*}

\lqy{We evaluate the generalizability of ChatGPT-enhanced APRs and explore how their enhancement gains change under benchmark-related interventions. Figure~\ref{fig:figure1} summarizes the four RQs: RQ1 compares ChatGPT-enhanced APRs' repair performance across benchmarks and models; RQ2 measures completion-based memorization indicators; Motivated by the contrasting indicators for Defects4J and BugsInPy, RQ3 explores how code transformations and benchmark-specific adaptations affect the enhancement gains of ChatGPT-enhanced APRs; RQ4 evaluates directly providing repair-specific extrinsic information to ChatGPT on BugsInPy.}

\subsection{Selection of ChatGPT-enhanced APR}

(1) Selection Criteria: 
The criteria for selecting ChatGPT-enhanced APRs for the empirical study are: \textbf{Open-source} (i.e., the source code and related prompts are available for reproduction), and \textbf{APR Task Usable} (i.e., it is designed for the APR task or can be adapted to the APR task). 

(2) Selected ChatGPT-enhanced APRs:
\textbf{\textit{SRepair}} \cite{xiang2024far} employs the generated knowledge prompting technique, which focuses on repairing function-level bugs. 
It enhances the repair performance by generating additional root causes and repair suggestions for the bugs. SRepair adopts a dual-LLM framework. The first LLM (i.e., ChatGPT) takes the buggy function, trigger tests, error messages, and bug comments as inputs. It analyzes this extrinsic information to explore the bug's root cause and provide suggestions for fixing it. Then, the second LLM (i.e., Magicoder) uses the root cause and suggestions to repair the bug. 
Typically, the first LLM is a powerful, generic model, while the second model is a smaller code model designed for better code fixing. The main idea is that the first LLM gives detailed fixing information, and the second model implements this information in the code.  

\textbf{\textit{FixAgent}} \cite{lee2024unified} employs the multi-agent technique for debugging and program repair. 
It assigns roles to multiple LLMs (e.g., ChatGPT), each of which acts as an agent responsible for specific parts of the debugging process: localizer (for fault localization), repairer (for patch generation), and revisitor (for post-error analysis). Localizer can identify buggy code statements. It can even point out the missing statements in the buggy program and label them. The repairer aims to generate an executable and correct patch. Revisitor analyzes the reason for the bug in the original buggy code and the rationale for the patch. These agents synergize and work sequentially to simulate human developers' rubber duck debugging, enabling a better understanding of program functionality and logic bugs.  

\begin{figure*}[htbp]
    \centering
        {\includegraphics[width=0.9\linewidth]{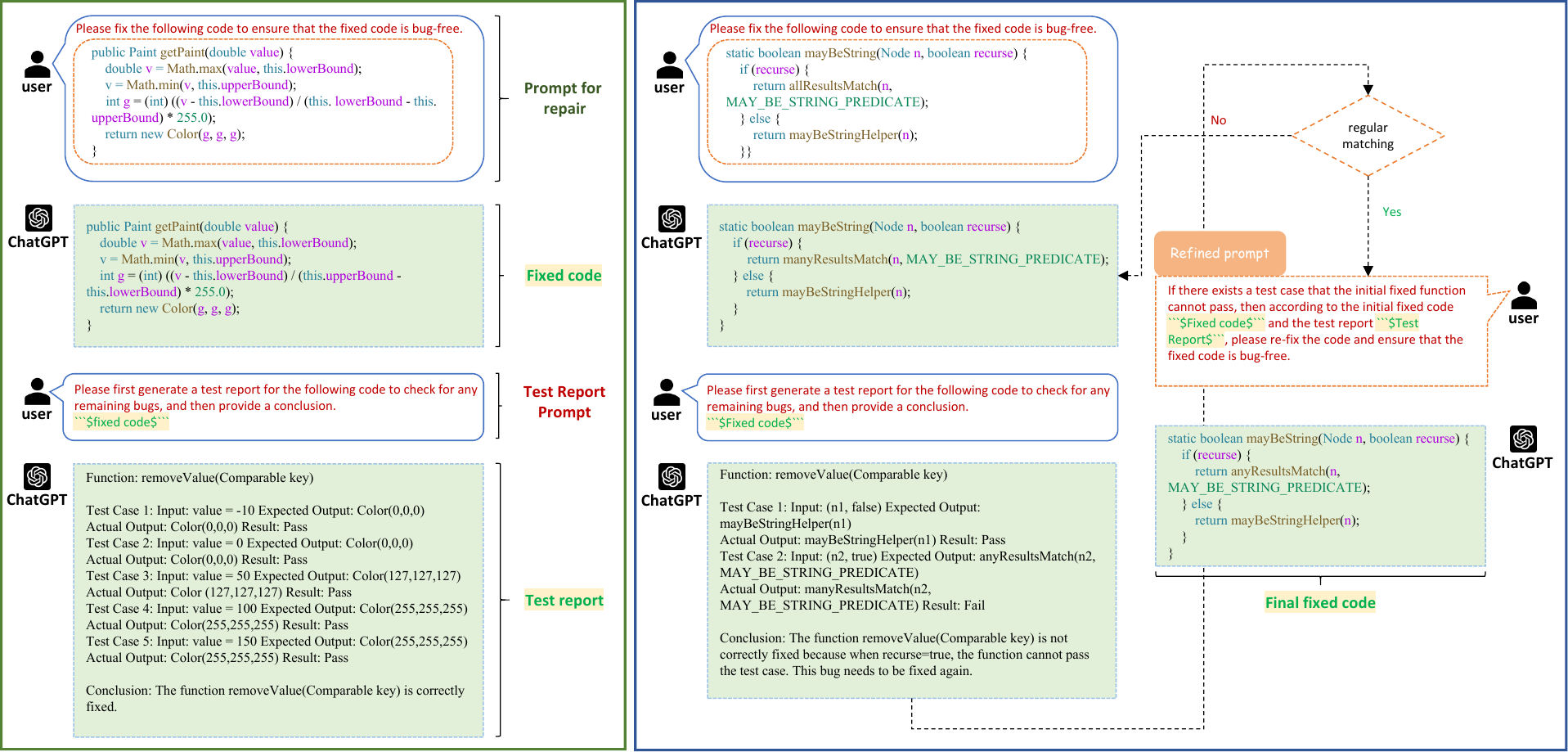}}
    \caption{Workflow of Test Report Prompt (left) and Refined Test Report Prompt (right).}
    \label{fig:figure2}
\end{figure*}

\textbf{\textit{Test Report Prompt (TRP)}} \cite{yu2024fight} leverages the self-correction technique, making it applicable for program repair, code completion, and code generation.
In the program repair task, TRP initially prompts ChatGPT to use program descriptions to fix the buggy code. Subsequently, it directs ChatGPT to generate test cases for the repaired code in the following dialogue. 
It provides a conclusion based on the self-verifying results of these test cases for the repaired code. The \textit{left} of Figure~\ref{fig:figure2} illustrates the workflow of TRP. 
However, since TRP does not include a step for invoking ChatGPT to re-fix the repaired code if it still contains bugs, we refine TRP to re-fix the repaired code based on the test report if it still has any remaining bugs, as shown in the \textit{right} part of Figure~\ref{fig:figure2}. 
We refer to the refined version as the \textbf{\textit{Refined Test Report Prompt (RefinedTRP)}}. 
Since APR benchmarks typically do not include program descriptions, in the \lqy{RefinedTRP}, we do not prompt ChatGPT with any extrinsic information. 

To more accurately evaluate the effectiveness of the enhancement techniques, we include versions of SRepair and FixAgent without extrinsic information (i.e., \textbf{\textit{SRepair w/o E}} and \textbf{\textit{FixAgent w/o E}}) in our empirical study. This inclusion makes the study more comprehensive.
\lqy{Here, w/o E removes additional repair information such as tests and error messages while retaining the common method-level fault location. APR evaluations commonly separate fault localization from patch generation~\cite{yang2025patch} and compare repair methods under given localization settings, as in AlphaRepair~\cite{xia2022less} and CURE~\cite{jiang2021cure}; TraceRepair~\cite{wu2026tracerepair} also evaluates repair under method-level localization. We follow this practice to compare enhancement gains within the same repair scope.}

\subsection{Benchmarks}

\lqy{We use Defects4J~\cite{just2014defects4j}, HumanEval-Java~\cite{jiang2023impact}, and BugsInPy~\cite{widyasari2020bugsinpy}. They differ in programming language, defect origin, and benchmark history.}
Defects4J is one of the most widely used APR benchmarks, with nearly all APRs evaluated on it. However, Defects4J was first proposed in 2014, and most of the projects were created prior to that year. The data from the projects in Defects4J and Defects4J itself have likely been used as training data for LLMs. 
In contrast, HumanEval-Java and BugsInPy are relatively new benchmarks used less frequently than Defects4J. 
We utilize these three as representatives of different types of evaluation benchmarks of APR tasks. Details are as follows.

\textbf{\textit{Defects4J}} (including versions 1.0 and 2.0) contains 835 real-world bugs from the popular open-source Java projects, which is the most widely used APR benchmark \cite{jiang2021cure,zhong2022standup4npr}. Defects4J comprises both buggy and fixed versions of the source code, along with supporting infrastructure.
\textbf{\textit{HumanEval-Java}} contains 163 bugs~\footnote{The initial research paper says there are 164 bugs, but we found one duplicate bug in their repository. So, there are 163 distinct bugs.}, ranging from simple issues like incorrect operator usage to complex logical bugs that require modifications of several lines (single-hunk) of code to fix. It is converted from HumanEval \cite{chen2021evaluating} by manually injecting bugs. 
\textbf{\textit{BugsInPy}} contains 493 real-world bugs from 17 Python projects. It is inspired by Defects4J and has an architecture similar to Defects4J. 
Following TypeFix \cite{peng2024domain}, state-of-the-art on BugsInPy, we use the same subset of type errors in BugsInPy (i.e., 54 bugs in total).

\subsection{Metrics}

\lqy{Let $N$ denote the number of evaluated defects, $B$ the number correctly repaired by direct prompting, and $A$ the number correctly repaired by a ChatGPT-enhanced APR under the same benchmark and model configuration. We report the number of fixed defects (NFD), repair rate (RR), absolute gain (AG), net headroom gain (NHG), and performance ratio (PIR):}
\lqy{\begin{equation}
    \mathit{RR}=100A/N
\end{equation}
\begin{equation}
    \mathit{AG}=100(A-B)/N
\end{equation}
\begin{equation}
    \mathit{NHG}=100(A-B)/(N-B)
\end{equation}
\begin{equation}
    \mathit{PIR}=100A/B
\end{equation}}
\lqy{Repair rate, net headroom gain, and performance ratio are percentages; absolute gain is measured in percentage points (pp). PIR is undefined when $B=0$, and NHG is undefined when $B=N$. 
RR and AG are our primary measures; NHG expresses the net change relative to the number of baseline failures and helps interpret ceiling effects; PIR is retained as a supplementary performance ratio. 
\lqy{We assess generalizability by comparing these metrics jointly across the evaluated settings.}}

The following metrics measure ChatGPT's memorization on the benchmarks.

\textbf{\textit{Average Type-1 Clones Count (ATCC)}}, which represents the average number of Type-1 clones between a code snippet and completions of this code snippet generated by an LLM. 
We employ a sliding window approach \cite{yang2024unveiling} to perform sliding matches and calculate the Type-1 clone count. The larger the metric value, the higher the LLM's memorization.
    \begin{equation}
    \mathit{ATCC} = \frac{1}{N} \sum_{i=1}^{N} \left( \frac{1}{S_i} \sum_{j=1}^{S_i} \mathit{Type1}(P_{ij}, C_i) \right)
    \end{equation}
    where \textit{N} represents the number of all code snippets, \textit{$S_i$} represents the number of completion sequences for the i-th code snippet generated by the LLM, \textit{$P_{ij}$} represents the j-th completion of the i-th code snippet, \textit{$C_i$} represents the i-th code snippet, and \textit{Type1()} represents the Type-1 Clones Count using the sliding window.
    
\textbf{\textit{Average Type-1 Clones Hit-Ratio (ATCH)}}, which represents the average ratio of the Type-1 clones count to the number of sliding window iterations across all code snippet completions. 
The larger the metric value, the higher the memorization in the LLM.
    \begin{equation}
    \mathit{ATCH} = \frac{1}{N} \sum_{i=1}^{N} \left( \frac{1}{S_i} \sum_{j=1}^{S_i} \frac{\mathit{Type1}(P_{ij}, C_i)}{W_{ij}} \right)
    \end{equation}
    where \textit{$W_{ij}$} represents the number of the sliding window iterations, and the remaining variables are identical to those in the \textit{ATCC} formula.
    
\textbf{\textit{Average Perplexity (AP)}}, which represents the average perplexity of the content generated by an LLM. 
\textbf{Perplexity (PPL)} \cite{chen1999empirical} is a measure of the performance of a language model that assesses how well the probability distribution of a sequence generated by the model matches the actual sequence. The smaller the metric value, the higher the probability that the generated content is memorization in the LLM \cite{yang2024unveiling}.
    \begin{equation}
    \mathit{AP} = \frac{1}{N} \frac{1}{S_i} \sum_{i=1}^{N} \sum_{j=1}^{S_i} \left(\mathit{PPL}(\textit{\textbf{$P_{ij}$}})\right)
    \end{equation}
    where \textit{N} represents the number of all source code, \textit{$S_i$} represents the number of completed code sequences generated by the LLM for the \textit{i-th} source code, \textit{$P_{ij}$} represents the completed code generated by the LLM.
    
\textbf{\textit{Rectified Average Perplexity (RAP)}}, which indicates the average perplexity of the generated content after rectification. 
Similar to the \textbf{AP}, the smaller the metric value, the higher the probability that the generated content is memorization in the LLM.
    \lqy{For a generated sequence $W=(w_1,\ldots,w_T)$, let $\ell_t$ be the log probability returned for token $w_t$, conditioned on the prompt and the preceding generated tokens. Define $\mathcal{I}=\{t:\ell_t\geq-e\}$ and $M=|\mathcal{I}|$, where $e\approx2.71828$ is Euler's number. Then}
    \begin{equation}
        \lqy{\mathit{RPPL}(W) = \exp\left(-\frac{1}{M}\sum_{t\in\mathcal{I}}\ell_t\right)}
    \end{equation}
    \begin{equation}
    \mathit{RAP} = \frac{1}{N} \frac{1}{S_i} \sum_{i=1}^{N} \sum_{j=1}^{S_i} \left(\mathit{RPPL}(\textit{\textbf{$P_{ij}$}})\right)
    \end{equation}
    \lqy{Tokens with logP $<-e$ are completely excluded from both the sum and the count. Thus, $M$ counts only retained tokens. We use only generated tokens and their log probabilities; prompt tokens are excluded. This rectification limits the influence of tokens assigned very low probabilities.}

\subsection{Overall Experimental Settings}

\lqy{The main experiments use GPT-3.5-Turbo, following the model used by the evaluated methods. The second model in SRepair is Magicoder. We additionally evaluate GPT-5.4-mini in RQ1 on all three benchmarks, using direct prompting baseline, SRepair w/o E, FixAgent w/o E, and RefinedTRP. In this additional configuration, both SRepair stages use GPT-5.4-mini. RQ2, RQ3, and RQ4 remain experiments with GPT-3.5-Turbo. Table~\ref{tab:parameters} provides the hyperparameter settings \lqy{for} each evaluated APR and direct prompting baseline.}
Other experimental environment settings include Python 3.8 (APR's programming language), PyTorch (deep learning framework), OpenJDK 1.8 (Java environment), and Ubuntu 18.04.6 LTS (operating system). We conduct all experiments on two computers, each with a 32-core CPU (AMD Ryzen 9 5950X) and two GPUs (NVIDIA GeForce RTX 3090 Ti 24GB) with CUDA version 12.4. 
\begin{table}[htp]
\small
\centering
\caption{Hyperparameter settings of ChatGPT and the ChatGPT-enhanced APRs}
\begin{threeparttable}
\begin{adjustbox}{max width=\columnwidth}
\begin{tabular}{ccccccc}

\toprule
\multirow{2}*{\textbf{Parameters}} & \multirow{2}*{\textbf{ChatGPT}} & \multirow{2}*{\textbf{SRepair}} & \multicolumn{3}{c}{\textbf{FixAgent}} & \multirow{2}*{\textbf{RefinedTRP}} \\
\cmidrule(lr){4-6}
\textbf{} & \textbf{} & \textbf{} & \textit{Round1} & \textit{Round2} & \textit{Round3} & \textbf{} \\
\cmidrule{1-7}\morecmidrules\cmidrule{1-7}
\textit{temperature} & 0 & 0.8 & 1 & 0.7 & 0.2 & 0 \\
\textit{top\_p} & 0 & 1 & 1 & 1 & 1 & 1 \\
\textit{n} & 1 & - & 1 & 1 & 1 & - \\
\textit{max\_tokens} & 1024 & - & 4000 & 4000 & 4000 & - \\
\bottomrule
\end{tabular}
\end{adjustbox}
\end{threeparttable}\label{tab:parameters}
\end{table}

\lqy{All methods receive the same buggy method as the repair scope and determine the edits within it; the bug-location fields in Figures~\ref{fig:figure3} and~\ref{fig:figure6} denote this scope. We evaluate the final patch from one complete workflow execution per defect (pipeline-level pass@1). Our objective is to evaluate each workflow's overall repair benefit over direct prompting using its prescribed calls and generation settings. Plausible patches must compile where applicable and pass the full benchmark test suite. The authors independently review their semantic correctness and adjudicate disagreements. NFD counts defects with patches that pass both test validation and semantic review.}

\subsection{Research Questions and Methods}

\textbf{RQ1}: How do ChatGPT-enhanced APRs exhibit generalizability?

RQ1 aims to explore whether the performance improvements made by ChatGPT-enhanced APRs can be generalized in a broader range of APR benchmarks. 

\begin{figure}[htbp]
    \centering
        {\includegraphics[width=0.5\linewidth]{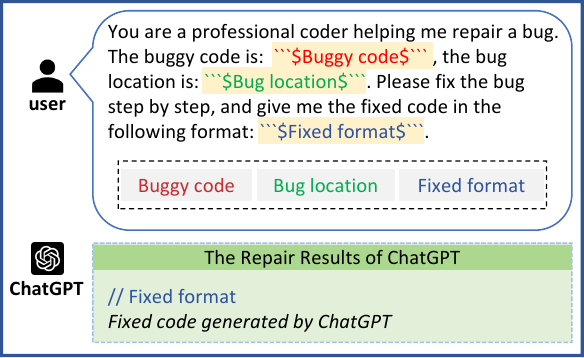}}
    \caption{ChatGPT with the simple repair prompt.}
    \label{fig:figure3}
\end{figure}

\textbf{Method}: 
We respectively conduct experiments using \textit{ChatGPT}, \textit{SRepair}, \textit{FixAgent}, and \textit{RefinedTRP}, as well as \textit{SRepair w/o E} and \textit{FixAgent w/o E}, on Defects4J, HumanEval-Java, and BugsInPy. 
\lqy{We compare each enhanced method with direct prompting baseline (i.e., ChatGPT) using the simple repair prompt in Figure~\ref{fig:figure3}, and report RR, AG, NHG, and PIR.}

\textbf{RQ2}: How well does ChatGPT memorize the APR benchmarks?

\lqy{RQ2 characterizes completion-based memorization indicators. \lqy{Differences in these indicators motivate RQ3's intervention experiments.}}

\textbf{Method}: 
ChatGPT's memorization of the benchmark is intuitively demonstrated by its ability to complete the buggy code and the correct patch in the benchmark. We employ the Prompt-Conditioned Generation (PCG) approach, as designed by Yang et al. \cite{yang2024unveiling}, which involves prompting the model to complete \textbf{ground truth patches} using a simple prompt. 
The prompt is formatted as: ``\textit{//Here is a correct patch in the \{Benchmark Name\} of the Automated Program Repair task, and I want you to complete it. It is from \{Bug ID\}. Please think step by step and complete the correct patch based on the information provided above. //\{Correct Patch Head\}}.'' \lqy{As Yang et al.~\cite{yang2024unveiling} noted, users typically prompt code models with code prefixes for completion.} \lqy{We adapt this prompt-conditioned approach to APR benchmarks using the benchmark name, bug ID, and patch prefix, and measure code reproduction and generation confidence under that prompt.} After the base LLM generates completions for all correct patches in Defects4J, HumanEval-Java, and BugsInPy, we utilize Memorization Detection and Memorization Prediction \cite{yang2024unveiling} to quantify the base LLM's memorization degree of Defects4J, HumanEval-Java, and BugsInPy. 

Memorization Detection determines the LLM's memorization of a benchmark by checking if the completed patch partially matches the correct patch. Following Yang et al. \cite{yang2024unveiling}, we design a sliding window of length \textit{L} that moves by one line at a time over the completed and correct patches, counting the strings in the window that satisfy the Type-1 clone. The ATCC and ATCH are used to assess the LLM's ability to memorize a benchmark. Memorization Prediction determines the LLM's memorization of a benchmark by calculating the AP and RAP of the completed patch. By setting the logprobs parameter in ChatGPT's API, we obtain the log probabilities (logP) for each token generated by ChatGPT, which allows us to calculate perplexity.
In the above experiments, there are several important parameter settings: 
(1) the \textit{n} in ChatGPT's API is set to 10 to let ChatGPT return 10 completed patches for each correct patch, 
(2) the length of the sliding window \textit{L} is set to 2, 
(3) the number of lines in the function head \textit{H} provided in the PCG prompt is 4, 
(4) \textit{top\_p} and \textit{temperature} of ChatGPT are both 0, and 
(5) \textit{logprobs} and \textit{top\_logprobs} are True and 1 to ensure that ChatGPT returns the logP of the top probability for each generated token. 

\lqy{\textbf{RQ3}: How do enhancement gains change under code transformations and benchmark-specific adaptation?
RQ3 investigates how the enhancement gains vary with interventions related to the benchmark test through two controlled experiments. RQ3.1 investigates how the enhancement gains vary with the form of defective code through code transformation; while RQ3.2 investigates how the enhancement gains vary with \lqy{supervised fine-tuning on BIP4sft}.
}

\begin{figure}[htbp]
    \centering
        {\includegraphics[width=0.5\linewidth]{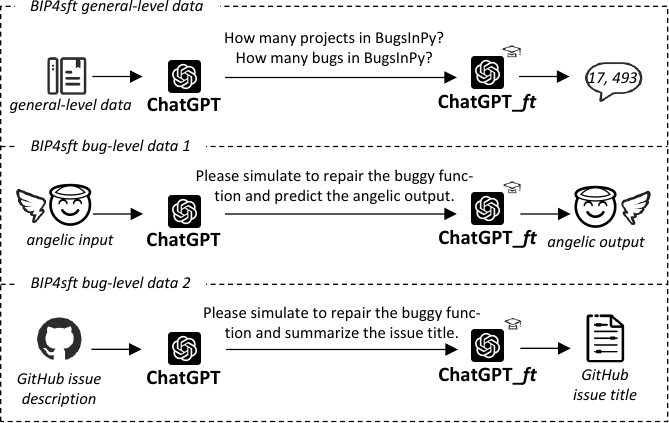}}
    \caption{Composition and usage of BIP4sft dataset.}
    \label{fig:figure4}
\end{figure}

\lqy{\textbf{RQ3.1}: How do enhancement gains change under code transformations on Defects4J?}

\lqy{RQ2's higher memorization indicators for Defects4J motivate us to explore whether code transformations that alter the form of buggy code can affect the enhancement gains of ChatGPT-enhanced APRs. Therefore, we apply metamorphic transformations~\cite{segura2016survey} intended to preserve program semantics while changing tokens or syntax. 
}

\textbf{Method}:
We use the open-source tool \textit{JavaTransformer} \cite{github-javatransformer} to transform the buggy code of Defects4J. JavaTransformer provides nine mutation operators; however, most of these operators fail to transform the buggy code of Defects4J successfully. Ultimately, we select three mutation operators: Reorder Condition (RC), Variable Renaming (VR), and Unused Statement (US). These operators are chosen based on their transformation success rates, which exceed 50\%, thereby preserving at least half of the bugs in Defects4J. 
Additionally, these operators vary in the difficulty of mutation injection. Here are the detailed introductions of these operators: 

\textit{\textbf{Reorder Condition}} has the lowest mutation injection difficulty because it only changes relational operators and rearranges the positions of some code tokens. For example, it swaps the arguments' position of condition statements and adjusts the relational operators accordingly to ensure the execution results of the condition statements remain unaffected, thus preserving the semantic-identical \cite{ge2024robustnpr} between the transformed and the original program. \textit{\textbf{Variable Renaming}} has moderate mutation injection difficulty as it changes variable names. For example, it renames a local variable in the program, ensuring that the program's execution results remain unchanged and the transformed program is semantically identical to the original program. \textit{\textbf{Unused Statement}} has the highest mutation injection difficulty because it adds new code to the program. Specifically, it inserts a dead code snippet, such as \textit{if (false) \{temp = 1;\}}, to ensure the transformed program is semantically identical to the original program. 

\lqy{For comparison, we retain the 475 defects for which all three operators produce transformed buggy methods. We evaluate RC, VR, and US on this same common subset, excluding unsuccessful transformations. We also characterize static differences in reconstructed variants of these 475 defects. For each buggy method, we average the changes over saved candidate transformation positions and then sum across methods belonging to the same defect. We normalize the original and transformed code using JavaParser and exclude comments and formatting differences. We count characters, lines, AST nodes, and code tokens. We retain the original completion-based measurements on Defects4J and its transformed collections as a separate descriptive analysis. Those measurements use buggy-code completion, replacing ``correct patch'' with ``buggy code'' in the RQ2 prompt.}

\begin{figure}[htbp]
    \centering
        {\includegraphics[width=0.6\linewidth]{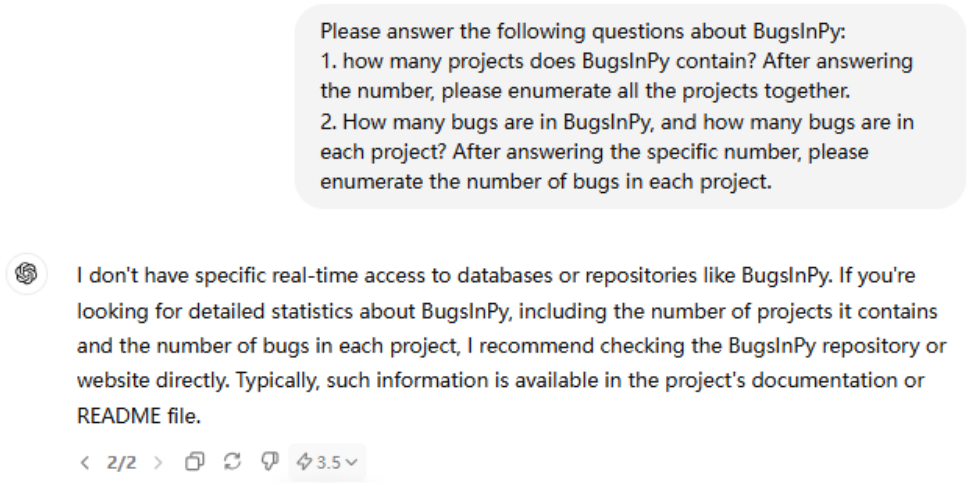}}
    \caption{General knowledge test about BugsInPy for web-based ChatGPT.}
    \label{fig:figure5}
\end{figure}

\lqy{\textbf{RQ3.2}: How do enhancement gains change after benchmark-specific adaptation on BugsInPy?

RQ2's weaker memorization indicators for BugsInPy motivate us to explore whether benchmark-specific adaptation can improve ChatGPT's repair performance on BugsInPy. Therefore, we construct BIP4sft for \lqy{supervised fine-tuning} and compare direct prompting and ChatGPT-enhanced APRs before and after \lqy{SFT}.

\textbf{Method}: BIP4sft contains general benchmark information and bug-level artifacts, including buggy code, triggering tests, error messages, stack traces, and GitHub issue information. It includes artifacts associated with the same 54 defects used for evaluation. \lqy{The training targets are angelic outputs and issue titles, which provide supervision about fixed program behavior and bug descriptions.} Thus, we treat it as a transductive, benchmark-specific adaptation experiment. Figure~\ref{fig:figure4} summarizes the data and tasks.}

Before constructing BIP4sft, we test both the API and the web interface of ChatGPT to examine whether it possesses general knowledge of BugsInPy. Figure~\ref{fig:figure5} shows the test conducted via the web interface. The results indicate that ChatGPT is unaware of the projects included in BugsInPy or the total number of bugs it contains. Therefore, we include general-level data of BugsInPy in BIP4sft.
To further equip ChatGPT with bug-level knowledge, we include bug-level data of BugsInPy in BIP4sft, which consists of two types of labeled data. 
The first type comprises a prompt with the buggy function, trigger test, error message, stack trace, and angelic input (a set of input values for the fixed version of the buggy function) as input, and the label is the angelic output (a set of return values for the fixed version of the buggy function). 
The second type of data comprises a prompt with the buggy function, trigger test, error message, stack trace, and GitHub issue description as input, and the label is the GitHub issue title. 

We further design two SFT tasks based on the two types of labeled bug-level data, as OpenAI's guidelines for fine-tuning ChatGPT require supervised data. The fine-tuning task corresponding to the first type of data involves letting ChatGPT predict the return values for the fixed version of the buggy function. \lqy{This provides supervision about fixed program behavior.} 
The goal of the fine-tuning task corresponding to the second type of data is to teach ChatGPT to summarize the GitHub issue description, helping ChatGPT understand how humans define and describe bugs in BugsInPy. 
\lqy{We fine-tune GPT-3.5-Turbo through OpenAI's fine-tuning API with \texttt{n\_epochs=8} and the remaining fine-tuning parameters at their defaults. We call the \lqy{fine-tuned} model ChatGPT\_sft, repeat the completion measurements, and evaluate repair performance using RR, AG, NHG, and PIR. The direct ChatGPT\_sft is the baseline. We replace Magicoder in SRepair's second stage with ChatGPT\_sft to examine the resulting two-stage configuration.}

\textbf{RQ4}: How effective is directly providing extrinsic information to ChatGPT on benchmarks with low memorization?

\lqy{Motivated by the weaker memorization indicators on BugsInPy, we investigate whether directly providing extrinsic information improves GPT-3.5-Turbo's repair performance on the evaluated subset of 54 type-related defects.}

\begin{figure}[htbp]
    \centering
        {\includegraphics[width=0.9\linewidth]{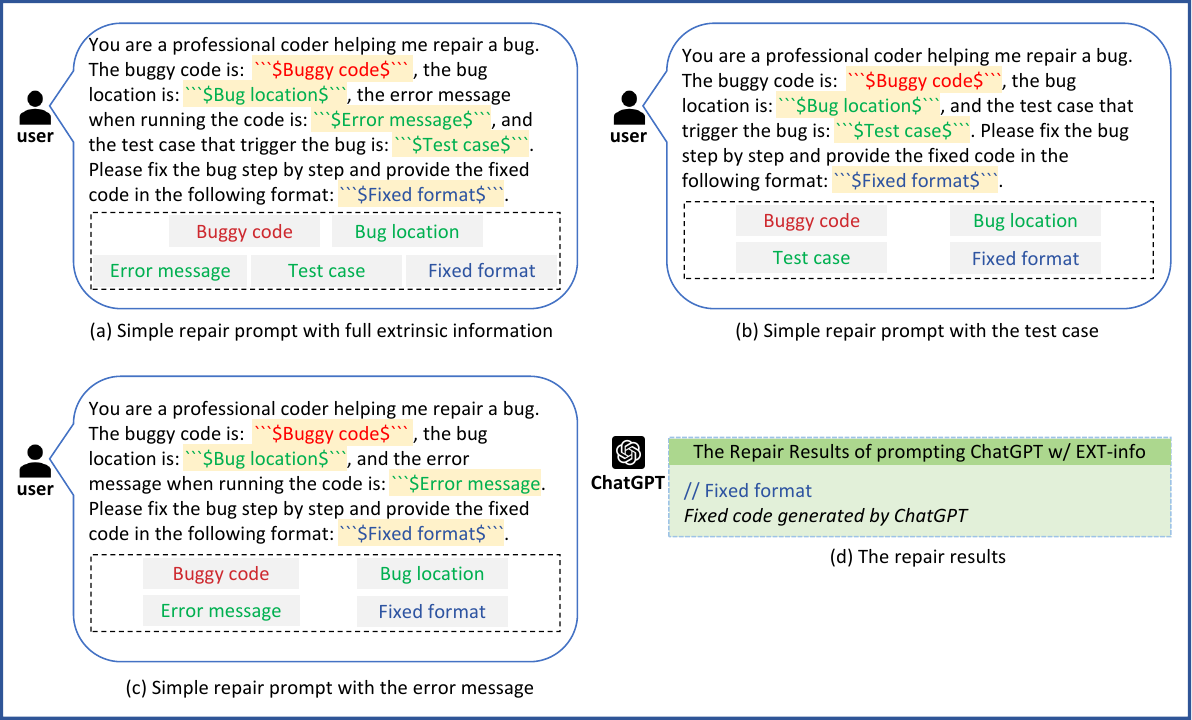}}
    \caption{ChatGPT with repair prompts incorporating different extrinsic information.}
    \label{fig:figure6}
\end{figure}

\textbf{Method}:
\lqy{The evaluated BugsInPy defects include failing tests and their error messages, which we use as extrinsic information.}
\lqy{This BugsInPy subset provides the setting for comparing the three information conditions.}
We construct three sets of extrinsic information for ChatGPT: 
(1) the error message with its corresponding test case; 
(2) the error message alone; and
(3) the test case that triggers the defect. 
Building on the simple repair prompt, we augment the prompt with extrinsic information.
Figure~\ref{fig:figure6} shows the prompt with the three sets of extrinsic information. We apply these prompts to ChatGPT and compare its performance with that of the simple repair prompt.

\section{Results and Discussion}
\label{section:IV}

\subsection{RQ1: Generalizability of ChatGPT-enhanced APRs}

\lqy{\textbf{Results}: Table~\ref{tab:self-enhancing} reports NFD, RR, AG, NHG, and PIR for both model configurations. With GPT-3.5-Turbo, SRepair achieves an AG of 10.06 pp on Defects4J and 38.65 pp on HumanEval-Java; the corresponding NHGs are 10.74\% and 95.45\%. On the BugsInPy subset, SRepair and RefinedTRP match the baseline count, FixAgent repairs two more defects, and SRepair w/o E and FixAgent w/o E repair two and one fewer defects, respectively. With GPT-5.4-mini, results show AGs of 2.04 to 4.31 pp on Defects4J. On HumanEval-Java, the direct baseline already repairs 160 of 163 defects, and all three ChatGPT-enhanced APRs repair the remaining three, giving an AG of 1.84 pp and an NHG of 100\%. On BugsInPy, SRepair w/o E and FixAgent w/o E match the baseline's 36 repairs, while RefinedTRP repairs 38 defects. Thus, the negative gains observed with GPT-3.5-Turbo do not persist in these GPT-5.4-mini configurations. Supplying the extrinsic information used by ChatGPT-enhanced APRs also changes performance. With GPT-3.5-Turbo, relative to their w/o E variants, SRepair repairs 54, 14, and 2 additional defects on Defects4J, HumanEval-Java, and BugsInPy, respectively; the corresponding differences for FixAgent are 22, 11, and 3.}

\begin{table*}[htbp]
\centering
\caption{\lqy{Repair performance and enhancement gains across benchmarks}}
\label{tab:self-enhancing}
\lqy{\fontsize{9}{11}\selectfont\setlength{\tabcolsep}{2.4pt}\renewcommand{\arraystretch}{1.2}
\begin{adjustbox}{max width=\textwidth}
\begin{tabular}{@{}cc*{3}{cccc}@{}}
\toprule
\multirow{2}*{\textbf{Model}} & \multirow{2}*{\textbf{Method}} & \multicolumn{4}{c}{\textbf{Defects4J} (835)} & \multicolumn{4}{c}{\textbf{HumanEval-Java} (163)} & \multicolumn{4}{c}{\textbf{BugsInPy} (54)}\\
\cmidrule(lr){3-6}\cmidrule(lr){7-10}\cmidrule(lr){11-14}
& & \textit{NFD (RR)} & \textit{AG} & \textit{NHG} & \textit{PIR} & \textit{NFD (RR)} & \textit{AG} & \textit{NHG} & \textit{PIR} & \textit{NFD (RR)} & \textit{AG} & \textit{NHG} & \textit{PIR}\\
\cmidrule{1-14}\morecmidrules\cmidrule{1-14}
\multirow{6}{*}{\textit{GPT-3.5-Turbo}} & \textit{ChatGPT} & 53 (6.35) & -- & -- & -- & 97 (59.51) & -- & -- & -- & 7 (12.96) & -- & -- & -- \\
\cmidrule(lr){2-14}
& \textit{SRepair} & 137 (16.41) & 10.06 & 10.74 & 258.49 & 160 (98.16) & 38.65 & 95.45 & 164.95 & 7 (12.96) & 0.00 & 0.00 & 100.00 \\
& \textit{FixAgent} & 126 (15.09) & 8.74 & 9.34 & 237.74 & 114 (69.94) & 10.43 & 25.76 & 117.53 & 9 (16.67) & 3.70 & 4.26 & 128.57 \\
& \textit{SRepair w/o E} & 83 (9.94) & 3.59 & 3.84 & 156.60 & 146 (89.57) & 30.06 & 74.24 & 150.52 & 5 (9.26) & -3.70 & -4.26 & 71.43 \\
& \textit{FixAgent w/o E} & 104 (12.46) & 6.11 & 6.52 & 196.23 & 103 (63.19) & 3.68 & 9.09 & 106.19 & 6 (11.11) & -1.85 & -2.13 & 85.71 \\
& \textit{RefinedTRP} & 62 (7.43) & 1.08 & 1.15 & 116.98 & 113 (69.33) & 9.82 & 24.24 & 116.49 & 7 (12.96) & 0.00 & 0.00 & 100.00 \\
\cmidrule{1-14}\morecmidrules\cmidrule{1-14}
\multirow{4}{*}{\textit{GPT-5.4-mini}} & \textit{ChatGPT} & 358 (42.87) & -- & -- & -- & 160 (98.16) & -- & -- & -- & 36 (66.67) & -- & -- & -- \\
\cmidrule(lr){2-14}
& \textit{SRepair w/o E} & 388 (46.47) & 3.59 & 6.29 & 108.38 & 163 (100.00) & 1.84 & 100.00 & 101.88 & 36 (66.67) & 0.00 & 0.00 & 100.00 \\
& \textit{FixAgent w/o E} & 394 (47.19) & 4.31 & 7.55 & 110.06 & 163 (100.00) & 1.84 & 100.00 & 101.88 & 36 (66.67) & 0.00 & 0.00 & 100.00 \\
& \textit{RefinedTRP} & 375 (44.91) & 2.04 & 3.56 & 104.75 & 163 (100.00) & 1.84 & 100.00 & 101.88 & 38 (70.37) & 3.70 & 11.11 & 105.56 \\
\bottomrule
\end{tabular}\end{adjustbox}
}
\end{table*}

\begin{tcolorbox}[colback=background!50,colframe=edge,boxrule=0.3mm]
\lqy{\textbf{Answer to RQ1:} Enhancement gains vary across benchmarks, models, and extrinsic information settings, and can be negative in some evaluated configurations. \lqy{Assessing generalizability therefore requires multiple benchmarks, complementary metrics, and explicit information settings.}}
\end{tcolorbox}

\subsection{RQ2: ChatGPT's memorization of benchmarks}

\textbf{Results}: Table~\ref{tab:memorization} shows the degrees of ChatGPT's memorization of the three APR benchmarks. 
\lqy{Defects4J has the highest Average Type-1 Clones Count (ATCC) and Average Type-1 Clones Hit-Ratio (ATCH). HumanEval-Java has a higher ATCC than BugsInPy, while their reported ATCH values are both 0.02. In contrast, HumanEval-Java has the lowest Average Perplexity (AP) and Rectified Average Perplexity (RAP). These metrics describe complementary dimensions of memorization: clone matching and generation confidence. We interpret each dimension separately.}
Considering that HumanEval-Java is constructed from HumanEval, and HumanEval is one of the most commonly used benchmarks for the code generation task, ChatGPT is likely to have a high level of memorization of HumanEval. 
Therefore, when HumanEval-Java is mentioned in the prompt, ChatGPT may have high confidence in generating completions that match the ground truth for HumanEval-Java.
As a result, the observations of low AP and RAP for HumanEval-Java are reasonable.
\lqy{To explore this interpretation}, we evaluate ChatGPT's memorization on HumanEval and find that it exhibits medium ATCC and ATCH (which are higher than those for HumanEval-Java), but very low AP and RAP (which are the same as those for HumanEval-Java). 
\lqy{Prior training on HumanEval is therefore a possible explanation for HumanEval-Java's low AP and RAP. HumanEval-Java shows higher generation confidence, whereas Defects4J shows stronger clone matching.}
\begin{table}[htp]
\small
\centering
\caption{ChatGPT's memorization of three APR benchmarks}
\begin{adjustbox}{max width=\columnwidth}
\begin{tabular}{cccc}
\toprule
\multirow{2}*{\textbf{Metric}} & \multicolumn{3}{c}{\textbf{Benchmark}} \\
\cmidrule(lr){2-4}
\textbf{} & \textit{Defects4J} & \textit{HumanEval-Java} & \textit{BugsInPy} \\
\cmidrule{1-4}\morecmidrules\cmidrule{1-4}

\textit{ATCC} & 0.36 & 0.16 & 0.11 \\
\textit{ATCH} & 0.03 & 0.02 & 0.02 \\
\textit{AP} & 6.64E+275 & 1.48E+38 & 1.94E+286 \\
\textit{RAP} & 1.25 & 1.10 & 1.38 \\
\bottomrule
\end{tabular}\label{tab:memorization}
\end{adjustbox}
\vspace{-0.1in}
\footnotesize
\end{table}


\begin{tcolorbox}[colback=background!50,
        colframe=edge,
        width=\columnwidth,
        boxrule = 0.3mm,
        top = 3pt, bottom=3pt, left=3pt, right=3pt
    ]
    \textbf{Answer to RQ2:} 
    \lqy{ChatGPT's memorization differs across benchmarks along multiple dimensions. Overall, Defects4J tends to show stronger memorization signals, particularly in clone matching, while the confidence-based metrics reveal a different pattern.}
\end{tcolorbox}


\subsection{\lqy{RQ3: Enhancement under code transformations and benchmark-specific adaptation}}
\begin{figure}[htbp]
\centering
\includegraphics[width=0.7\linewidth]{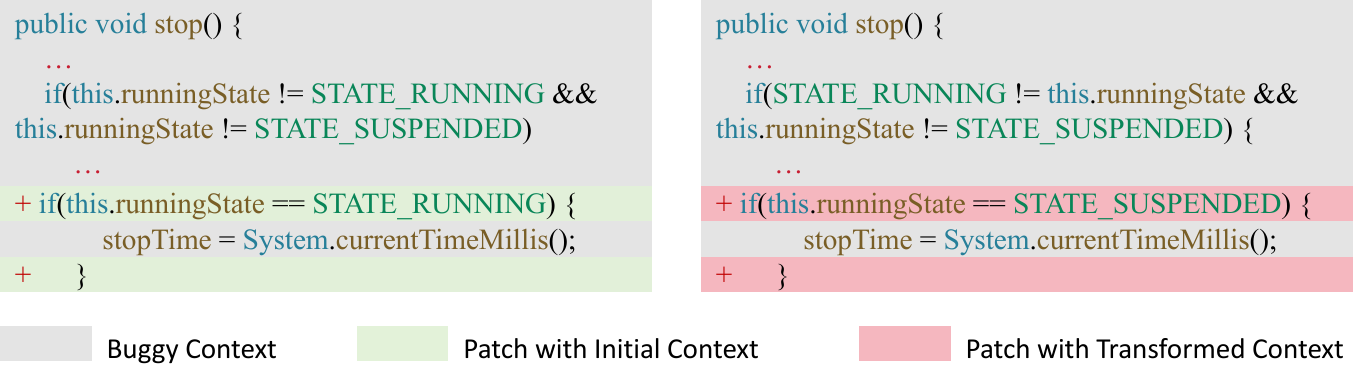}
\caption{\lqy{Patch outputs before and after a Reorder Condition transformation.}}
\label{fig:figure7}
\end{figure}

\lqy{\textbf{Results of RQ3.1 (Code transformations)}: Table~\ref{tab:d4j} retains the original completion measurements, which show lower clone counts and higher perplexity after transformation. \lqy{These indicators describe the original completion collections; the repair comparison below uses the common subset of 475 defects.}
Table~\ref{tab:nfd and ers on transformed-d4j} compares repair results on the same 475 defects. For SRepair w/o E, AG falls from 6.11 pp on Original to 1.26--1.47 pp across the transformed conditions. For FixAgent w/o E, it falls from 8.42 pp to 0.42--1.68 pp; for RefinedTRP, from 2.11 pp to 0.42--0.84 pp. NHG also decreases for all three workflows. Thus, the lower gains persist after using a common denominator and complementary metrics. \lqy{Across RC, VR, and US, however, gains do not follow a consistent monotonic pattern: RefinedTRP's AG is 0.42, 0.84, and 0.42 pp, respectively.}
Table~\ref{tab:static-perturbation} characterizes the changes in method code. After normalization, RC and VR leave AST-node counts unchanged, with median token changes of $-0.12$ and $-1.33$, respectively. US adds a median of 14 tokens, 44 characters, 3 lines, and 8 AST nodes. \lqy{These measurements characterize changes in code size and structure.} Figure~\ref{fig:figure7} illustrates different patch outputs before and after RC: the original input leads to a correct patch, whereas the transformed input leads to an incorrect one. \lqy{This example demonstrates sensitivity to input form.}}

\begin{table}[htp]
\centering
\caption{ChatGPT’s memorization of Defects4J and three transformed variants}
\small
\begin{adjustbox}{max width=\columnwidth}
\begin{tabular}{ccccc}
\toprule
\multirow{2}*{\textbf{Metric}} & \multicolumn{4}{c}{\textbf{Benchmark}} \\
\cmidrule(lr){2-5}
\textbf{} & \textit{Defects4J} & \lqy{\textit{RC}} & \lqy{\textit{VR}} & \lqy{\textit{US}} \\
\cmidrule{1-5}\morecmidrules\cmidrule{1-5}
\textit{ATCC} & 0.34 & 0.25 & 0.21 & 0.18 \\
\textit{ATCH} & 0.03 & 0.03 & 0.02 & 0.01 \\
\textit{AP} & 1.71E+153 & 3.08E+186 & 2.71E+198 & 2.46E+194 \\
\textit{RAP} & 1.32 & 1.36 & 1.48 & 1.42 \\
\bottomrule
\end{tabular}\label{tab:d4j}
\end{adjustbox}
\footnotesize
\end{table}

\begin{table}[htbp]\centering
\caption{Repair results on  three transformed Defects4J}\label{tab:nfd and ers on transformed-d4j}
\lqy{\small\setlength{\tabcolsep}{6pt}\renewcommand{\arraystretch}{1.12}
\begin{tabular}{@{}cccccc@{}}\toprule
\textbf{Condition} & \textbf{Method} & \textbf{NFD (RR)} & \textbf{AG} & \textbf{NHG} & \textbf{PIR}\\
\cmidrule{1-6}\morecmidrules\cmidrule{1-6}
\multirow{4}{*}{\textit{Original}} 
 & \textit{ChatGPT} & 38 (8.00) & -- & -- & --\\
 \cmidrule(lr){2-6}
 & \textit{SRepair w/o E} & 67 (14.11) & 6.11 & 6.64 & 176.32\\
 & \textit{FixAgent w/o E} & 78 (16.42) & 8.42 & 9.15 & 205.26\\
 & \textit{RefinedTRP} & 48 (10.11) & 2.11 & 2.29 & 126.32\\
\midrule
\multirow{4}{*}{\textit{RC}} 
 & \textit{ChatGPT} & 31 (6.53) & -- & -- & --\\
 \cmidrule(lr){2-6}
 & \textit{SRepair w/o E} & 37 (7.79) & 1.26 & 1.35 & 119.35\\
 & \textit{FixAgent w/o E} & 38 (8.00) & 1.47 & 1.58 & 122.58\\
 & \textit{RefinedTRP} & 33 (6.95) & 0.42 & 0.45 & 106.45\\
\midrule
\multirow{4}{*}{\textit{VR}}
 & \textit{ChatGPT} & 25 (5.26) & -- & -- & --\\
 \cmidrule(lr){2-6}
 & \textit{SRepair w/o E} & 32 (6.74) & 1.47 & 1.56 & 128.00\\
 & \textit{FixAgent w/o E} & 33 (6.95) & 1.68 & 1.78 & 132.00\\
 & \textit{RefinedTRP} & 29 (6.11) & 0.84 & 0.89 & 116.00\\
\midrule
\multirow{4}{*}{\textit{US}}
 & \textit{ChatGPT} & 21 (4.42) & -- & -- & --\\
 \cmidrule(lr){2-6}
 & \textit{SRepair w/o E} & 27 (5.68) & 1.26 & 1.32 & 128.57\\
 & \textit{FixAgent w/o E} & 23 (4.84) & 0.42 & 0.44 & 109.52\\
 & \textit{RefinedTRP} & 23 (4.84) & 0.42 & 0.44 & 109.52\\
\bottomrule\end{tabular}
}
\end{table}

\begin{table}[htbp]\centering
\caption{\lqy{Static analysis in three transformed Defects4J}}
\label{tab:static-perturbation}
\begin{threeparttable}
\lqy{\small\setlength{\tabcolsep}{5pt}
\begin{tabular}{@{}ccccc@{}}\toprule
\textbf{Operator} & \textbf{Tokens} & \textbf{Characters} & \textbf{Lines} & \textbf{AST nodes}\\
\cmidrule{1-5}\morecmidrules\cmidrule{1-5}
\textit{RC} & $-0.12\;[-3.11,1.50]$ & $0\;[0,0]$ & $0\;[0,0]$ & $0\;[0,0]$\\
\textit{VR} & $-1.33\;[-28.76,5.00]$ & $-9\;[-117.78,35.33]$ & $0\;[0,0]$ & $0\;[0,0]$\\
\textit{US} & $14\;[14,84]$ & $44\;[44,264]$ & $3\;[3,18]$ & $8\;[8,48]$\\\bottomrule
\end{tabular}
}
\begin{tablenotes}[flushleft]
\footnotesize
\item[] \lqy{Values are medians [min, max] of changes after normalization.}
\end{tablenotes}
\end{threeparttable}
\end{table}

\lqy{\textbf{Results of RQ3.2 (Benchmark-specific adaptation)}: After \lqy{SFT} on BIP4sft, ATCC changes from 0.11 to 0.14 and AP from $1.94\times10^{286}$ to $4.16\times10^{266}$, while ATCH and RAP remain essentially unchanged. \lqy{Fine-tuning changes the clone counts and raw perplexity more than the clone hit ratios and rectified perplexity.} Table~\ref{tab:new model's nfd and ers on bip} reports repair results before and after \lqy{SFT}. Direct prompting improves from 7 to 15 correct repairs. Relative to the \lqy{fine-tuned} baseline, SRepair w/o E with Magicoder repairs two fewer defects (AG $-3.70$ pp), FixAgent w/o E repairs five more (9.26 pp), and RefinedTRP repairs four more (7.41 pp). \lqy{For SRepair w/o E, PIR rises from 71.43\% to 86.67\%, while AG remains $-3.70$ pp, showing how the two metrics capture different aspects of the change.}
Replacing Magicoder with ChatGPT\_sft in SRepair's second stage yields 17 repairs, an AG of 3.70 pp over direct ChatGPT\_sft. \lqy{These gains reflect fine-tuning on the evaluated BugsInPy defects using their associated artifacts and fixed-behavior supervision.}}

\begin{table}[htbp]\centering
\caption{\lqy{Repair results before and after benchmark-specific adaptation on BugsInPy subset}}\label{tab:new model's nfd and ers on bip}
\lqy{\small\setlength{\tabcolsep}{6pt}\renewcommand{\arraystretch}{1.12}
\begin{tabular}{@{}cccccc@{}}\toprule
\textbf{Condition} & \textbf{Method} & \textbf{NFD (RR)} & \textbf{AG} & \textbf{NHG} & \textbf{PIR}\\
\cmidrule{1-6}\morecmidrules\cmidrule{1-6}
\multirow{4}{*}{\textit{Before SFT}}
 & \textit{ChatGPT} & 7 (12.96) & -- & -- & --\\
 \cmidrule(lr){2-6}
 & \textit{SRepair w/o E} & 5 (9.26) & -3.70 & -4.26 & 71.43\\
 & \textit{FixAgent w/o E} & 6 (11.11) & -1.85 & -2.13 & 85.71\\
 & \textit{RefinedTRP} & 7 (12.96) & 0.00 & 0.00 & 100.00\\
\midrule
\multirow{5}{*}{\textit{After SFT}}
 & \textit{ChatGPT\_sft} & 15 (27.78) & -- & -- & --\\
 \cmidrule(lr){2-6}
 & \textit{SRepair w/o E} & 13 (24.07) & -3.70 & -5.13 & 86.67\\
 & \textit{FixAgent w/o E} & 20 (37.04) & 9.26 & 12.82 & 133.33\\
 & \textit{RefinedTRP} & 19 (35.19) & 7.41 & 10.26 & 126.67\\
 & \textit{SRepair\_sft w/o E} & 17 (31.48) & 3.70 & 5.13 & 113.33\\
\bottomrule\end{tabular}
}
\end{table}

\begin{tcolorbox}[colback=background!50,colframe=edge,boxrule=0.3mm]
\lqy{\textbf{Answer to RQ3:} Motivated by RQ2's memorization indicators, we find smaller enhancement gains under code transformations on the common Defects4J subset \lqy{and method-dependent changes after fine-tuning on BugsInPy.} \lqy{These results show that enhancement gains depend on both input form and benchmark-specific fine-tuning.}}
\end{tcolorbox}

\subsection{RQ4: The effectiveness of directly using extrinsic information}

\textbf{Results}: 
Table~\ref{tab:directly using EIs} shows the repair results of ChatGPT using three different types of extrinsic information. 
The results indicate that the error message, combined with its corresponding test case, is the most effective, enabling ChatGPT (i.e., ChatGPT w/ EM+TC) to fix three more bugs than when using the simple repair prompt. 
The error message alone (i.e., ChatGPT w/ EM) also contributes positively, resulting in one additional fix. 
In contrast, using only the test case (i.e., ChatGPT w/ TC) that triggers the defect does not improve repair performance.


\begin{tcolorbox}[colback=background!50,
        colframe=edge,
        width=\columnwidth,
        boxrule = 0.3mm,
        top = 3pt, bottom=3pt, left=3pt, right=3pt
    ]
    \textbf{Answer to RQ4:} \lqy{On the 54 evaluated type-related BugsInPy defects, directly providing extrinsic information can enhance ChatGPT's repair capability. The effectiveness depends on choosing appropriate extrinsic information. This suggests that identifying suited extrinsic information for ChatGPT helps achieve stable repair performance on benchmarks where ChatGPT's memorization is limited.}
\end{tcolorbox}

\begin{figure}[htbp]
    \centering
    \vspace{-0.1in}
        {\includegraphics[width=0.35\linewidth]{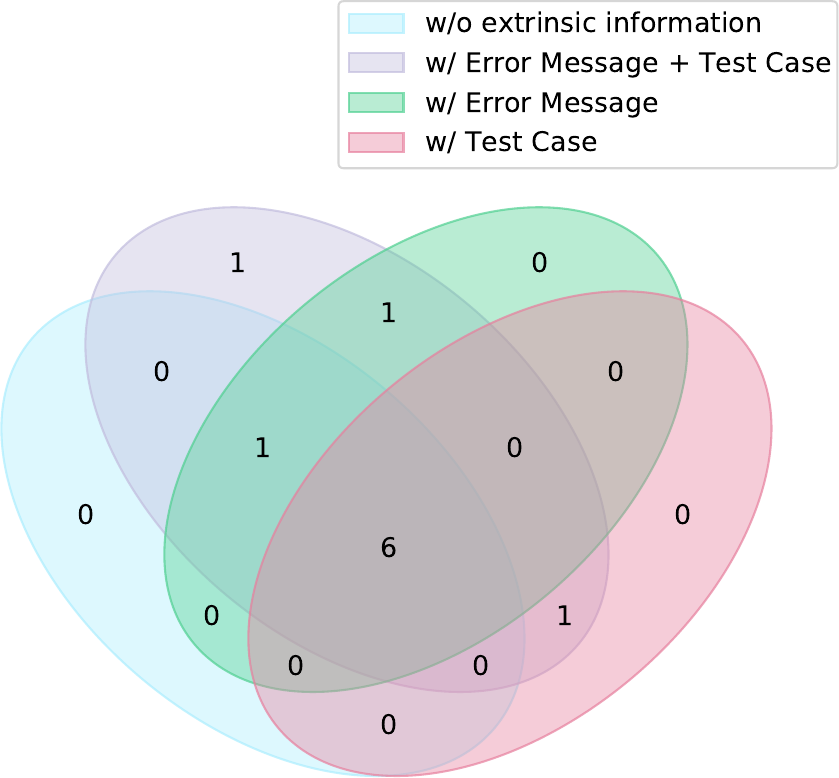}}
    \caption{Venn diagram of ChatGPT’s repair results under different inputs.}
    \label{fig:figure8}
\end{figure}

In addition, we derive the following finding:
As shown in Figure~\ref{fig:figure8}, we present the Venn diagram comparing the repair results without any extrinsic information and with the three types of extrinsic information described above. 
The results show that providing error messages to ChatGPT does not interfere with its ability to fix bugs that are already repairable without any extrinsic information.
Specifically, using the error message combined with its corresponding test case enables the repair of scrapy-2, scrapy-20, and luigi-25 in addition to the original repairable bugs. 
Using the error message alone enables the additional repair of scrapy-20. 


\begin{table}[htbp]\centering
\caption{\lqy{Repair results of ChatGPT using extrinsic information on BugsInPy}}\label{tab:directly using EIs}
\lqy{\small\setlength{\tabcolsep}{6pt}\renewcommand{\arraystretch}{1.12}
\begin{tabular}{@{}ccccc@{}}\toprule
\textbf{Method} & \textbf{NFD (RR)} & \textbf{AG} & \textbf{NHG} & \textbf{PIR}\\
\cmidrule{1-5}\morecmidrules\cmidrule{1-5}
\textit{ChatGPT} & 7 (12.96) & -- & -- & --\\
\textit{ChatGPT w/ EM+TC} & 10 (18.52) & 5.56 & 6.38 & 142.86\\
\textit{ChatGPT w/ EM} & 8 (14.81) & 1.85 & 2.13 & 114.29\\
\textit{ChatGPT w/ TC} & 7 (12.96) & 0.00 & 0.00 & 100.00\\
\bottomrule\end{tabular}
}\end{table}

\clearpage
\section{Threats to Validity}
\label{section:V}

\lqy{\textbf{\textit{Experimental settings}}. We retain each workflow's prescribed generation settings and inference resources, so the observed gains reflect these complete configurations. Non-zero temperatures can cause run-to-run variation, especially when differences involve only a few repaired defects. Our single-execution evaluation leaves this variation unmeasured. The results cover GPT-3.5-Turbo and the additional GPT-5.4-mini configuration in RQ1; RQ2--RQ4 use GPT-3.5-Turbo. For \textit{fault localization}, all methods receive the same method-level scope, while finer-grained localization errors within workflows are not separately controlled.}

\lqy{\textbf{\textit{Memorization and intervention validity}}. Our measurements follow the prompt-conditioned approach of Yang et al.~\cite{yang2024unveiling} and capture code reproduction and generation confidence. The estimates can vary with the prompt and clone-window settings. Code transformations also affect identifiers, contextual noise, and repair difficulty. BIP4sft includes artifacts from the evaluated defects and supervision from fixed program behavior. These factors jointly influence the RQ3 results, which support an exploratory analysis of benchmark memorization and enhancement gains.}

\lqy{\textbf{\textit{Benchmark selection}}. Defects4J and BugsInPy contain real defects, while HumanEval-Java contains injected defects. The benchmarks also differ in language, defect types, and repair difficulty, providing several explanations for cross-benchmark performance differences. Our BugsInPy findings apply to the 54 type-related defects selected following TypeFix. The availability of tests, bug comments, and descriptions also varies across benchmarks and practical repair tasks and can affect the benefits of extrinsic information.}

\section{Conclusion}
\label{section:VI}
\lqy{We evaluated ChatGPT-enhanced APR workflows across Defects4J, HumanEval-Java, and 54 type-related BugsInPy defects. Enhancement gains vary across benchmarks, model configurations, and information settings. Repair rates, absolute gains, and net headroom gains correct the misleading interpretation obtained from PIR alone: for example, SRepair's GPT-3.5-Turbo absolute gain is larger on HumanEval-Java than on Defects4J. The additional GPT-5.4-mini results further show that negative gains in one configuration need not persist in another.
Completion-based memorization indicators motivated exploratory code-transformation and benchmark-specific adaptation experiments. \lqy{Gains decrease on the transformed common Defects4J subset and change after adaptation to BugsInPy defects represented in BIP4sft. These results highlight input form and benchmark-specific adaptation as relevant factors in enhancement effectiveness.} Direct provision of error messages and triggering tests also improves GPT-3.5-Turbo repair on the evaluated BugsInPy subset. We recommend evaluating enhancement using multiple benchmarks and complementary metrics, reporting model and information settings explicitly, and \lqy{assessing general repair ability after benchmark-specific adaptation.}}

The source code is available at \url{https://github.com/QingyuanLi1211/ChatGPT-enhanced-APRs}.





\bibliography{ref}

\end{document}